\documentclass[
reprint,
superscriptaddress,
amsmath,
amssymb,
aps,
longbibliography,
prx 
]{revtex4-2}            
\usepackage{amsmath}    
\usepackage{mathtools}  
\usepackage{amssymb}    
\usepackage{graphicx}   
\usepackage{bm}         
\usepackage[colorlinks, linkcolor=blue, citecolor=blue, urlcolor=blue,breaklinks=true]{hyperref}
\usepackage{color,dsfont,multirow,booktabs}
\usepackage{braket}
\usepackage{tabularx}
\usepackage{lipsum,booktabs}
\usepackage{multirow}
\usepackage{float}

\begin{document}
\preprint{APS/123-QED}
\title{Topological Polarization Beam Splitter with Polarization-Selective Edge States}

\author{Shirin Afzal}
\affiliation{Institute for Quantum Science and Technology, and Department of Physics and Astronomy, University of Calgary,
2500 University Drive NW, Calgary, Alberta T2N 1N4, Canada}
\author{Amesh Kahloon}
\affiliation{Institute for Quantum Science and Technology, and Department of Physics and Astronomy, University of Calgary,
2500 University Drive NW, Calgary, Alberta T2N 1N4, Canada}
\author{Shabir Barzanjeh}
\email{Corresponding Author: shabir.barzanjeh@ucalgary.ca}
\affiliation{Institute for Quantum Science and Technology, and Department of Physics and Astronomy, University of Calgary,
2500 University Drive NW, Calgary, Alberta T2N 1N4, Canada}


\begin{abstract}
The realization of on-chip polarization beam splitters robust to fabrication imperfections remains a key challenge for polarization-sensitive photonic integration. We demonstrate a topologically protected polarization beam splitter based on a Floquet-engineered microring lattice implemented on a CMOS-compatible silicon nitride platform. By tailoring the dispersion of inter-ring coupling, the lattice supports complementary trivial and topological band gaps for orthogonal eigenpolarizations.
At telecom wavelengths, TE modes propagate via a topological edge state while TM modes are suppressed by a trivial gap; this behavior reverses at shorter wavelengths. We measure extinction ratios of 16–20 dB for the protected port and 10–20 dB for the non-protected port, with insertion loss of 2 dB at long wavelengths. Reduced TM extinction at shorter wavelengths is attributed to suboptimal input coupling.
We further identify spectral regions where both polarizations exhibit nontrivial band gaps, enabling polarization-independent edge transport and establishing a Floquet dual-polarization topological regime. Because operation is governed by band topology rather than geometric fine-tuning, the device shows intrinsic robustness to defects. These results establish polarization-tailored topological lattices as a scalable platform for robust beam splitting, routing, and interconnects in classical and quantum photonic systems.

\end{abstract}

\maketitle


\section*{Introduction}
Integrated photonics has emerged as a leading frontier in modern optics, concentrating complex information processing into chip-scale platforms. Its versatility spans quantum computing \cite{kok2007linear,wang2020integrated}, frequency-comb generation \cite{herr2016dissipative,chang2022integrated}, plasmonics \cite{dionne2010silicon,sorger2012toward}, and optomechanics \cite{safavi2019controlling,barzanjeh2022optomechanics}, where on-chip integration yields stability, scalability, and cost advantages that are difficult to match with bulk optics.

Progress in ultralow-loss waveguides and resonators across multiple material platforms has moved the field beyond isolated proofs of
concept to complex, deeply integrated circuits \cite{ Perez-Lopez2025}. As systems grow, fabrication nonidealities—sidewall roughness, thickness variations, and lithographic errors—accumulate, leading to scattering loss, phase slip, and polarization mismatch. Even on mature platforms, maintaining tight polarization specifications across broad bandwidths and large die areas remains challenging. These realities motivate architectures that deliver polarization-resolved functionality while remaining stable under realistic process variations.

Topological photonic insulators (TPIs) address part of this problem by supporting edge states protected by nontrivial band topology.
Chip-scale TPIs have enabled lasing \cite{bandres2018topological,nasari2023non}, entanglement generation \cite{mittal2021tunable,dai2022topologically,afzal2024enhanced}, and frequency-comb formation \cite{flower2024observation}. Their central advantage is the emergence of edge states—set by topological invariants—that are resilient to classes of local disorder and common fabrication defects \cite{lu2016topological,ozawa2019topological}. However, most demonstrations have not treated polarization as an explicit degree of freedom in the topology, leaving open opportunities for polarization-selective and polarization-independent transport \cite{hu2023observation}, including circular far-field polarization-dependent edge states \cite{parappurath2020direct} and topologically protected polarization beam splitters (PBS). A PBS, in general, routes Transverse Electric (TE) and Transverse Magnetic (TM) to distinct output ports at a common wavelength and is fundamental to multiplexing \cite{liu2019arbitrarily}, sensing \cite{bag2020towards}, and quantum information processing \cite{tanzilli2005photonic}. Non-topological PBS designs—Multimode Interferometers \cite{hong2003design,sun2017realization,zhan2021silicon}, photonic-crystal–assisted couplers \cite{shi2010experimental,xu2019compact,xu2021broadband}, directional couplers \cite{fukuda2006ultrasmall,li2017compact,kim2018high}, and inverse-designed devices \cite{shen2015integrated,xu2024inverse}—can achieve excellent figures of merit but typically require tight fabrication tolerances, stack-specific dispersion, and careful athermalization. In particular, photonic-crystal PBSs exploit polarization-selective band gaps \cite{shi2010experimental,xu2019compact, schonbrun2006polarization, zabelin2007self}, making wafer-scale integration sensitive to nanoscale disorder, etch bias, and coupling constraints. A topology-based approach that assigns polarization to protected edge channels enables robustness to local disorder, relaxed process margins, and flexible routing, opening new modes of polarization control on chip. 

\begin{figure*}[t!]
    \centering
    \includegraphics[width=1\linewidth]{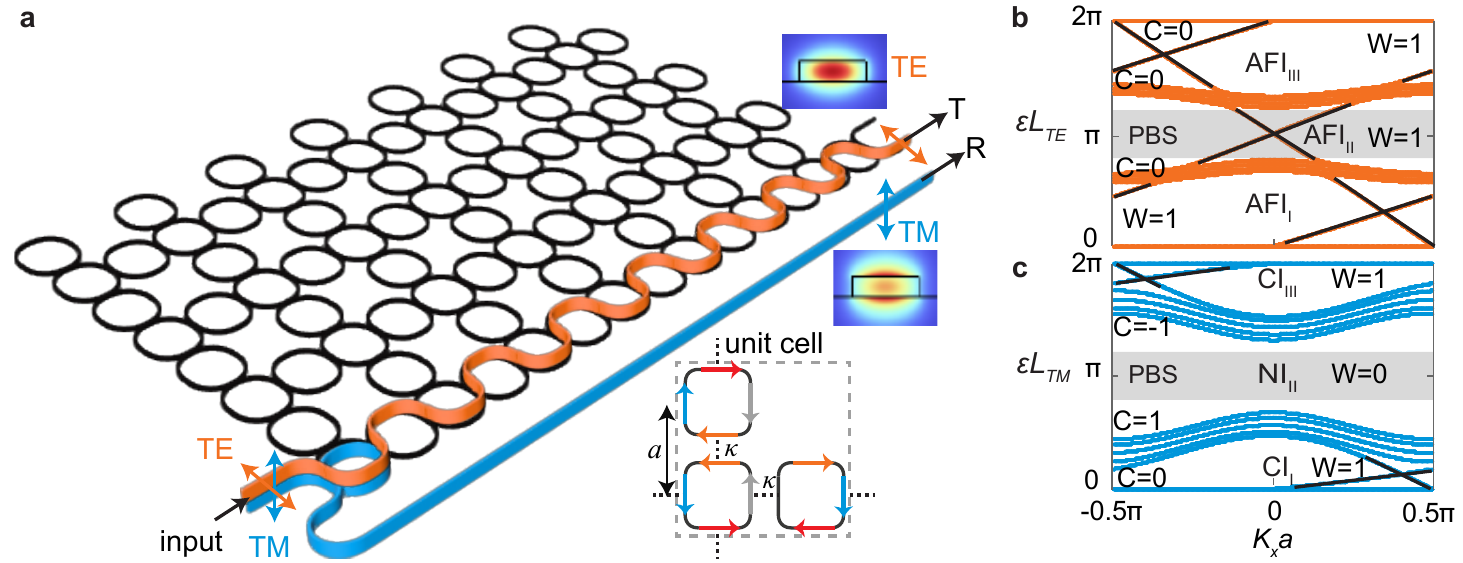}
    \caption{\textbf{Theory of a topological PBS.} \textbf{a} Schematic of a TE-protected topological PBS; the unit cell shown below contains three ring resonators characterized by coupling gap \(\kappa\) and center-to-center spacing \(a\). The orange trace depicts TE-polarized light guided by a topological edge state, while the blue trace indicates TM-polarized light reflected by a trivial (conventional) band gap. The output-port modes are obtained from simulations of the fundamental TE and TM modes with effective indices 1.687 and 1.596, respectively, at \(\lambda=1550\)~nm. \textbf{b,c} Simulated energy-band diagrams for TE and TM, respectively, for a semi-infinite lattice with four unit cells in the \(y\)-direction and an infinite number in \(x\). The coupling powers are \(\kappa_\mathrm{TE}^2=98.5\%\) (\(\theta_\mathrm{TE}=0.46\pi\)) and \(\kappa_\mathrm{TM}^2=53\%\) (\(\theta_\mathrm{TM}=0.26\pi\)). The lattice realizes an anomalous Floquet–insulator (AFI) phase for TE-polarized light, whereas TM-polarized light exhibits a Floquet–Chern–insulator (CI) phase separated by a normal-insulator (NI) gap. The topological invariants—the Chern (C) and the winding (W) numbers—are computed numerically. The gray region highlightsquasi-energies where lattice perform as PBS.
    }
    \label{fig:schematic}
\end{figure*}

Coupled microring TPIs \cite{afzal2018topological,afzal2020realization} realize Floquet-type (quasi-energy) bands \cite{kitagawa2010topological,maczewsky2017observation, rudner2013anomalous}  via engineered coupling between microrings
on a CMOS-compatible, low-loss platform, enabling broadband dispersion control and polarization-dependent topology, which opens trivial and nontrivial band gaps for TE and TM across multiple free-spectral ranges (FSRs). Building on this, we demonstrate—to our knowledge—the first Floquet topological PBS implemented on a single two-dimensional microring lattice. Dispersion engineering induces polarization-dependent band topology, producing edge states that selectively guide TE or TM modes with robust operation over multiple FSRs. Material dispersion sets a deterministic wavelength mapping: at shorter wavelengths, TM occupies the protected edge state, and TE is rejected, whereas at longer wavelengths, TE occupies the edge state, and TM is excluded. Because this mapping repeats for each FSR, polarization-selective routing persists over a broad spectral span without active tuning. In addition, at certain wavelengths, both TE and TM experience nontrivial band gaps simultaneously, yielding polarization-independent edge transport. In this architecture, polarization separation occurs at the input via bandgap filtering, while topological protection governs the subsequent propagation of the transmitted polarization by suppressing backscattering along the lattice boundary. The benefit is therefore not the act of splitting itself, but the stability of the transmitted channel and preservation of polarization contrast in the presence of disorder.

The device is fabricated from silicon nitride (SiN), which offers ultralow propagation loss, negligible two-photon absorption at telecom wavelengths, and CMOS-compatible processing, enabling low insertion loss and practical co-integration with nonlinear and quantum photonic circuits. Experimentally, we measure polarization extinction ratios of 10–16 dB sustained across entire FSRs in the C and L bands. Each FSR is $~1.4$ nm wide, with up to $50\%$ of an FSR operating as a PBS window. The measured insertion loss is 2 dB for TE and 6 dB for TM, primarily due to coupling mismatch at the bus–lattice interface, rather than disorder-induced scattering within the lattice.

Our results establish a novel polarization-management strategy that combines fabrication tolerance with broadband, FSR-periodic operation in a single lattice. By tying band topology to polarization, the architecture provides an independent control knob for routing, splitting, and combining light in the photonic circuits. The same design principles extend to polarization routers, beam combiners, and interconnects resilient to polarization-mode dispersion, with direct relevance to polarization-encoded quantum photonics and broadband \(\chi^{(3)}\) platforms. Rather than a single PBS, this approach defines a platform for polarization-programmable, topologically robust building blocks for integrated photonics.

\section*{RESULTS}
\subsection*{Device concept and lattice design}

\begin{figure*}[t!]
    \centering
    \includegraphics[width=1\linewidth]{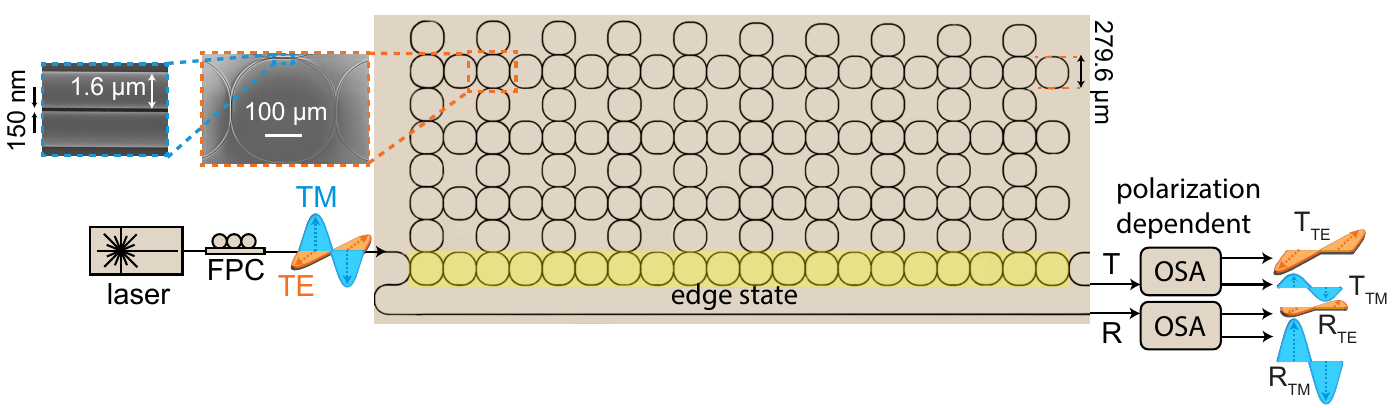}
    \caption{\textbf{Experimental setup and device micrographs}. Optical micrograph of the fabricated microring-lattice TPI (\(4\times10\) unit cells) on a SiN-on-insulator platform (400~nm SiN with 4.5~\(\mu\)m lower and 3~\(\mu\)m upper SiO\(_2\) cladding). The lattice comprises square microrings formed from 1.6~\(\mu\)m-wide waveguides with 150~nm coupling gap. Insets show SEM images of the coupled waveguides and a representative square microring with a side length of 38 $\mu$m and corners with radius of 120~\(\mu\)m. For characterization, a tunable continuous-wave (CW) laser and a fiber-based polarization controller (FPC) set the input polarization. Light is edge-coupled into a bus waveguide on the left. The portion of light that bypasses coupling into the lattice, together with light reflected by the trivial band gap, is collected at port \textbf{R}; light transmitted through the TPI is collected at port \textbf{T}. A polarization-resolved optical spectrum analyzer (OSA) separates and records TE and TM-polarized lights at both \textbf{T} and \textbf{R}.}
    \label{fig:experimental_setup}
\end{figure*} 
Figure~\ref{fig:schematic}\textbf{a} shows the TPI microring lattice that implements the PBS, a two-dimensional array of coupled square rings with one input port on the left and two output ports on the right, labeled \textbf{T} (through/transmission) and \textbf{R} (reflection). For measurement convenience, the reflection path is re-routed so that both ports are collected at the same chip facet. The unit cell (inset) comprises three evanescently coupled microrings with a uniform coupling coefficient \(\kappa\) and center-to-center spacing \(a\). Light circulating in each ring couples periodically to its nearest neighbors with a coupling angle \(\theta\), defined via the power transfer relation \(\kappa^{2}=\sin^{2}\!\theta\). The periodic coupling steps are shown in Fig.~\ref{fig:schematic}\textbf{a} by arrows in different segments of microrings within a unit cell. Because \(\theta\) depends on material dispersion and optical polarization, the lattice can be dispersion-engineered so that the two linear polarizations (TE and TM) experience different effective couplings over a chosen spectral window.
One representative scenario is illustrated in Figs.~\ref{fig:schematic}\textbf{b,c}, which show the band structures in a single Floquet–Brillouin zone for a semi-infinite strip (five unit cells along \(y\), periodic in \(x\)). Over the frequency range of interest, the coupling is strong for TE, \(\kappa_{\mathrm{TE}}^{2}=98.5\%\) \((\theta_{\mathrm{TE}}=0.46\pi)\), and weaker for TM, \(\kappa_{\mathrm{TM}}^{2}=53\%\) \((\theta_{\mathrm{TM}}=0.26\pi)\). Note that the topological invariants (Chern (C) and winding (W) numbers) of the gaps are computed numerically, following Ref.~\cite{afzal2018topological}.

\begin{figure*}
    \centering
    \includegraphics[width=1\linewidth]{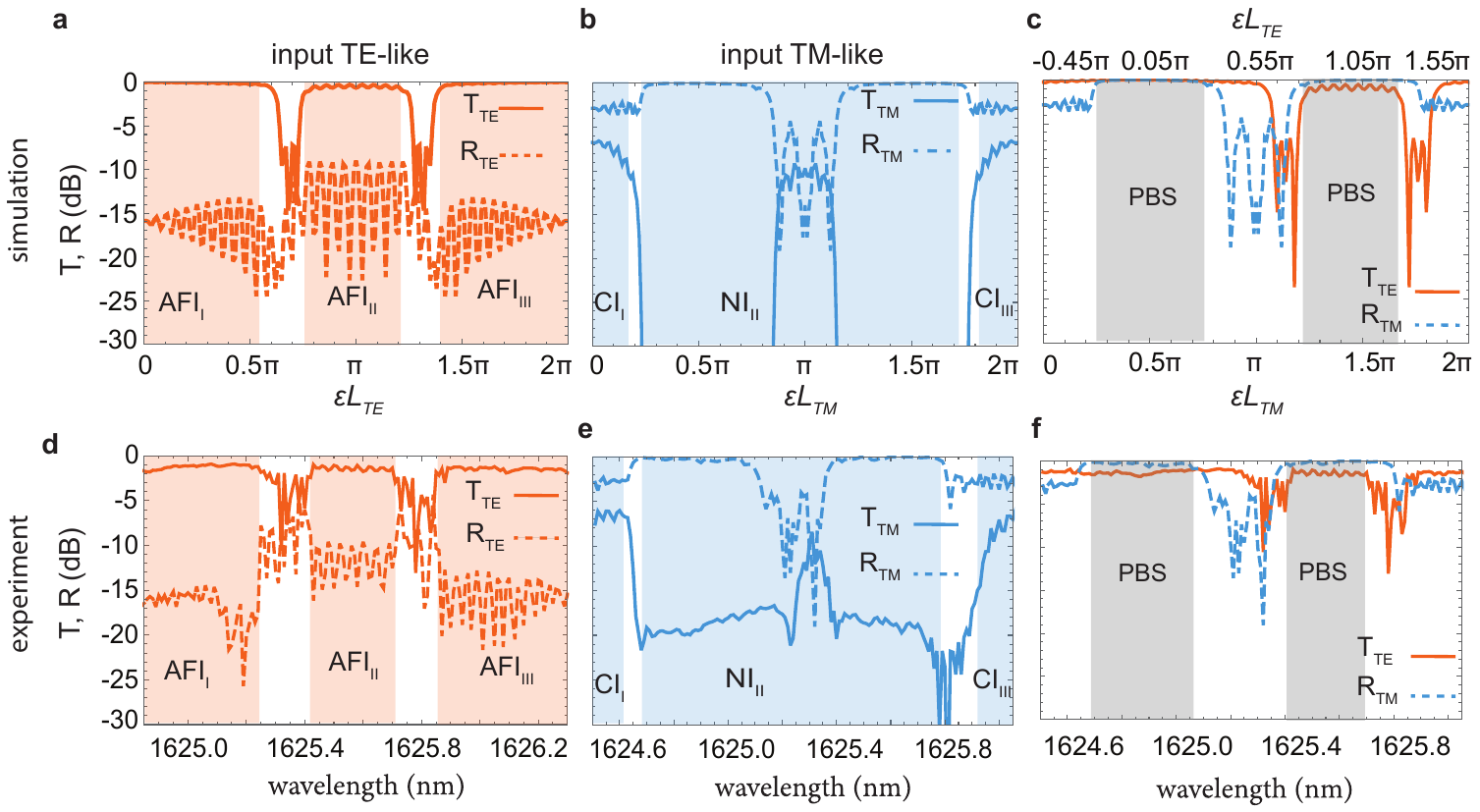}
    \caption{\textbf{TE-protected topological PBS.} \textbf{a}, \textbf{b} depict the simulated intensity of TE- and TM-polarized light at port \textbf{T} (solid line trace) and port \textbf{R} (dashed line trace). These simulations use coupling powers of $\kappa_\text{{TE}}^2 = 98.5 \%$ ($\theta_\text{{TE}} = 0.46\pi$) for TE-polarized light (orange trace) and $\kappa_\text{{TM}}^2 = 53 \%$ ($\theta_\text{{TM}} = 0.74\pi$) for TM-polarized light (blue trace) respectively. 
    \textbf{c} Simulated TE and TM-polarized light from the output ports of the PBS (\textbf{T} and \textbf{r}) in one combined TE and TM FSR while considering the phase offset between TE- and TM-polarized light. The round-trip loss for both TE and TM estimated from measurements is  $a_{rt}=0.975$. \textbf{d} and \textbf{e} Measured power from the \textbf{T} and \textbf{R} ports for TE-like and TM-like polarized light respectively. Orange and blue traces show the TE and TM-polarized parts of the measured power respectively. \textbf{f} Measured TE and TM-polarized light from the ports (\textbf{T} and \textbf{r}) of the PBS. The gray area indicates the wavelengths where the lattice works as a TE-protected PBS. }
    \label{fig:Experiment_TE_PBS}
\end{figure*}

For this coupling regime, the lattice realizes an anomalous Floquet–insulator (AFI) phase for TE-polarized light (Fig.~\ref{fig:schematic}\textbf{b}), while TM-polarized light exhibits a Floquet–Chern–insulator (CI) phase separated by normal-insulator (NI) gaps (Fig.~\ref{fig:schematic}\textbf{c}). The corresponding edge states (black lines in Fig.~\ref{fig:schematic}\textbf{b}, \textbf{c}) show that TE-polarized light with quasi-energy in the central band gap propagates along the lattice edge to port \textbf{T}, whereas TM at the same frequency lies in a NI gap and is rejected to \textbf{R} (see Fig.~\ref{fig:schematic}\textbf{a}). The gray band in Fig.~\ref{fig:schematic}\textbf{b},\textbf{c} shows the spectral window over which the device functions as a PBS. By exploiting material dispersion, the coupling hierarchy can be reversed at selected wavelengths so that
\(
\theta_{\mathrm{TM}}>\theta_{\mathrm{TE}}
\) (see Methods). In this regime, the TM-polarized light is topologically protected, while the TE-polarized light is reflected by the lattice. This wavelength-dependent assignment enables the polarization-selective edge transport used in our PBS. In this work, protection refers specifically to the suppression of backscattering during the propagation of the selected polarization along the lattice boundary.

We fabricated a microring-lattice TPI comprising \(4\times10\) unit cells on a SiN-on-insulator platform (400 nm SiN with 4.5 \(\mu\)m lower and 3 \(\mu\)m upper SiO\(_2\) cladding). The lattice consists of square microrings formed from 1.6~\(\mu\)m-wide waveguides with a 150~nm gap between adjacent microrings.
To mitigate corner loss associated with the relatively low effective index (\(\sim\!1.7\)), we rounded the ring corners to a 120~\(\mu\)m radius, which increased the effective coupling length for both TE and TM
polarized light. Numerical mode analysis was used to extract the coupling strength per unit length, \(\theta/\ell_{\mathrm{TE/TM}}\). We selected a side length of 38~\(\mu\)m to engineer polarization-dependent coupling dispersion, yielding TM-protected topological PBS behavior at shorter wavelengths and TE-protected behavior at longer wavelengths (see Methods).

\subsection*{Experiment}

An optical micrograph of the fabricated lattice is shown in Fig.~\ref{fig:experimental_setup}; insets show the SEM images of the coupled waveguides and a representative square microring. For characterization, a tunable continuous-wave (CW) laser and a fiber-based polarization controller (FPC) set the input polarization. Light is edge-coupled into a bus waveguide on the left side of the lattice, with an input--lattice coupling coefficient matched to the coupling between adjacent microrings.
Light that fails to couple from the bus into the lattice, together with light rejected by the trivial (normal) band gap, exits at the reflection port \textbf{R},
whereas light that couples into and propagates through the lattice exits at the through port \textbf{T}. 
A polarization-resolved optical spectrum analyzer (OSA) separates and records the TE and TM spectra at both \textbf{T} and \textbf{R} ports.

To evaluate the PBS performance, we conducted two measurement sets: one around a higher-wavelength FSR centered near 1625~nm and another around a lower-wavelength FSR near 1515~nm. Figure~\ref{fig:Experiment_TE_PBS} compares simulations and measurements at the longer-wavelength FSR for TE-like and TM-like inputs. Simulations were carried out using analytical methods~\cite{Tsay2011}, assuming identical microring round-trip loss for TE and TM $a_{\mathrm{rt}}=0.975$. Note that the FSRs for TE and TM are not identical due to the polarization-dependent dispersion of the effective indices.

For TE-like input, the coupling power between microrings is estimated as \(\kappa_{\mathrm{TE}}^{2}\approx98.5\%\),  corresponding to a coupling angle \(\theta_{\mathrm{TE}}=\sin^{-1}(\kappa_{\mathrm{TE}})\approx0.46\pi\) (extracted from Fig.~\ref{fig:coupling_sim}). For TM-polarized light, the coupling power is \(\kappa_{\mathrm{TM}}^{2}=53\%\), giving \(\theta_{\mathrm{TM}}=\pi-\sin^{-1}(\kappa_{\mathrm{TM}})=0.74\pi\).

Figures~\ref{fig:Experiment_TE_PBS}\textbf{a} and \textbf{b} show the simulated normalized transmission (solid) and reflection (dashed) for both TE- and TM-polarized light. In Fig.~\ref{fig:Experiment_TE_PBS}\textbf{a}, three plateau regions align with the \(\mathrm{AFI}_{\mathrm{I,II,III}}\) band gaps (orange shading). The orange dashed line indicates TE-polarized light that is reflected and has failed to couple into the lattice due to coupling mismatch or back-reflection. In Fig.~\ref{fig:Experiment_TE_PBS}\textbf{b}, in the normal band gap near mid-FSR, the lattice reflects essentially all incident TM-polarized light (blue dashed curve). A narrow transmission window appears inside this band gap (blue solid curve), reducing the reflected power; we attribute this feature to surface states that arise when the coupling angle exceeds \(\pi/2\). A detailed analysis of these states is beyond the scope of this work. The simulations also predict CI phases for TM-polarized light in the first and third band gaps.

Figures~\ref{fig:Experiment_TE_PBS}\textbf{d} and \textbf{e} show the measured TE (orange) and TM (blue) spectra at the transmitted (\textbf{T}, solid) and reflected (\textbf{R}, dashed) ports. The TE data agree well with the simulations. By contrast, the TM signal sits at the noise floor at \textbf{T} and appears only at \textbf{R}, indicating TE-dominated transmission and near-complete reflection of TM from the lattice.

\begin{figure*}
    \centering
    \includegraphics[width=1\linewidth]{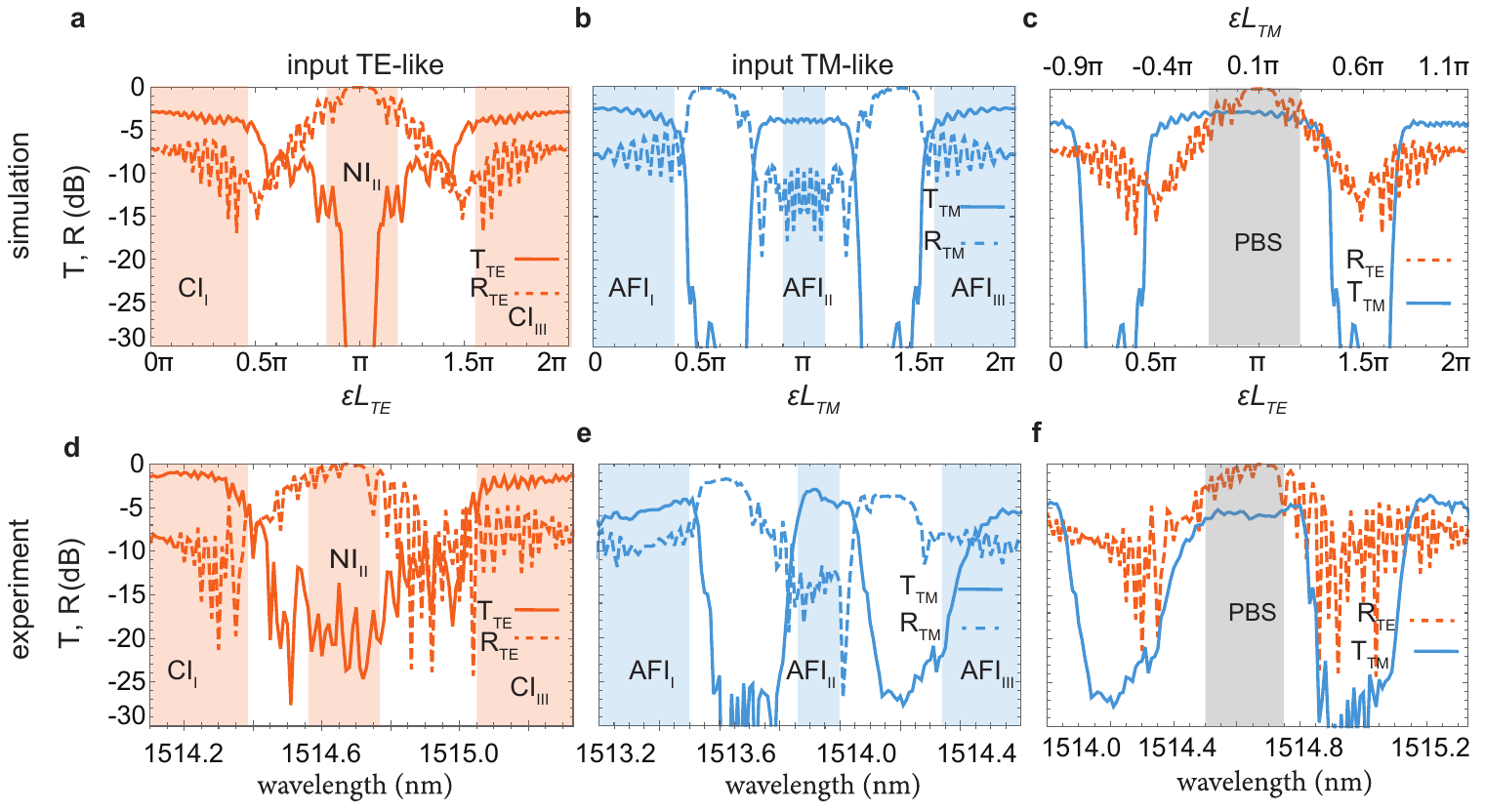}
    \caption{\textbf{TM-protected topological PBS.} \textbf{a}, \textbf{b} Simulated intensity of TE and TM-polarized light at port T, solid-line trace, and R, dashed-line trace,  using the coupling power of
    $\kappa_\text{{TE}}^2=82\%$ ($\theta_\text{{TE}}=0.36\pi$) and $\kappa_\text{{TM}}^2=84\%$ ($\theta_\text{{TM}}=0.63\pi$), respectively for the TE (orange trace) and TM (blue trace) lights. \textbf{c} Simulated TE- and TM-polarized light from output ports of PBS (T and R) in one combined TE and TM FSR considering the phase offset between TE- and TM-polarized light. \textbf{d} and \textbf{e} The measured power, respectively, from ports \textbf{T} and \textbf{R}  for TE-like and TM-like polarized light. Orange and blue traces respectively show the TE and TM-polarized parts of measured power. \textbf{f} Measured TE and TM-polarized light from output ports (\textbf{T} and \textbf{R}) of PBS. The gray area highlights the wavelengths where our lattice works as a TM-protected PBS.}
    \label{fig:Experiment_TM_PBS}
\end{figure*}

Comparing simulation and experiment reveals an approximate \(0.46\pi\) phase offset in a single FSR between the TE- and TM-polarized responses; this TE–TM phase offset varies slightly across different FSRs. Figure~\ref{fig:Experiment_TE_PBS}\textbf{c} illustrates the resulting behavior: in spectral regions where TE-polarized light propagates along the lattice edge under topological protection, TM-polarized light is fully reflected by the normal band gap. This arises because spectral regions that are topologically nontrivial for one polarization are simultaneously topologically trivial for the other. This complementary behavior routes TE and TM to different output ports and defines the PBS operating windows, highlighted by the gray regions in Fig.~\ref{fig:Experiment_TE_PBS}\textbf{c}. The measurements in Fig.~\ref{fig:Experiment_TE_PBS}\textbf{f} confirm these PBS windows: for the present device, the polarization extinction ratios $\text{PER}_\text{TE}=10\,\text{log}_{10}(\text{\textbf{T}}_{\text{TE}}/\text{\textbf{T}}_{\text{TM}})$ and $\text{PER}_\text{TM}=10\,\text{log}_{10}(\text{\textbf{R}}_{\text{TM}}/\text{\textbf{R}}_{\text{TE}})$ are 16-20 dB and 10-20 dB for TE- and TM-polarized light, respectively.

As explained above, the same lattice can also function as a TPI-based PBS for TM-polarized light. Simulations (Fig.~\ref{fig:Experiment_TM_PBS}\textbf{a},\textbf{b}) show that in one FSR, the TE response exhibits two CI band gaps and one NI band gap, whereas the TM response exhibits three AFI band gaps. These behaviors correspond to coupling powers \(\kappa_{\mathrm{TE}}^{2}\!\approx\!82\%\) (yielding \(\theta_{\mathrm{TE}}=\sin^{-1}(\kappa_{\mathrm{TE}})=0.36\pi\)) and \(\kappa_{\mathrm{TM}}^{2}\!\approx\!84\%\) (yielding \(\theta_{\mathrm{TM}}=\pi-\sin^{-1}(\kappa_{\mathrm{TM}})=0.63\pi\)). The measured spectra (Fig.~\ref{fig:Experiment_TM_PBS}\textbf{d},\textbf{e}) agree well with these predictions for TE and TM, respectively. Comparing simulation and experiment, we infer a TE–TM phase offset of approximately \(0.9\pi\); the simulated phase evolution is shown in Fig.~\ref{fig:Experiment_TM_PBS}\textbf{c}. The experimental results for TE- and TM-polarized light (Fig.~\ref{fig:Experiment_TM_PBS}\textbf{f}) confirm that the lattice
operates as a PBS, with TM as the topologically protected channel. However, weak coupling between the bus waveguides and the lattice
reduces the transmitted power at port~\textbf{T} for TM.
Because a fraction of TM light does not couple into the lattice, it appears at the reflection port~\textbf{R}, which, in turn, lowers
the measured extinction. The resulting PERs are approximately 7-25 dB and 7-20 dB for  TE and TM polarized light, respectively.

\begin{figure*}
    \centering
   \includegraphics[width=1\linewidth]{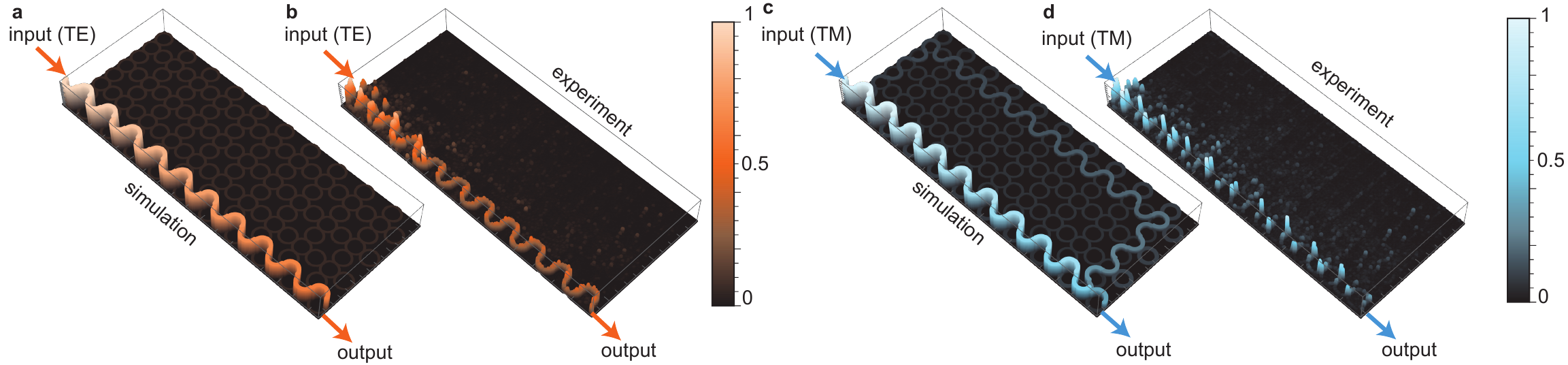}
    \caption{\textbf{.} \textbf{a} and \textbf{b} simulation and experimental results of the PBS at wavelength 1624.85 nm where the PBS topologically protect TE-polarized light. \textbf{c} and \textbf{d} simulation and experimental result of the PBS at the lower wavelength 1514.7 nm, where TM mode is topologically protected.  }
    \label{fig:NIR_image}
\end{figure*}

To visualize the topologically protected PBS and the emergence of TE and TM-polarized edge states, Fig.~\ref{fig:NIR_image} presents simulations alongside near-infrared images at a longer wavelength (\(\lambda = 1624.85~\mathrm{nm}\)) and a shorter wavelength (\(\lambda = 1514.7~\mathrm{nm}\)). 
Propagation is modeled with an analytical coupled-microring framework in which TE/TM enter through the coupling power between microrings \(\kappa^{2}\) and the round-trip quasi-energy \(\epsilon L_{\mathrm{TE/TM}}\) of a single ring.

For the longer-wavelength case (Fig.~\ref{fig:NIR_image}\textbf{a}), the simulation uses \(\kappa_{\mathrm{TE}}^{2}=98.5\%\) and \(\kappa_{\mathrm{TM}}^{2}=53\%\) with \(\epsilon L_{\mathrm{TE}}=0.05\pi\) and \(\epsilon L_{\mathrm{TM}}=0.5\pi\). For the shorter-wavelength case (Fig.~\ref{fig:NIR_image}\textbf{c}), we use \(\kappa_{\mathrm{TM}}^{2}=84\%\) and \(\kappa_{\mathrm{TE}}^{2}=82\%\) with \(\epsilon L_{\mathrm{TM}}=0.1\pi\) and \(\epsilon L_{\mathrm{TE}}=\pi\). These parameter values are taken from the gray operating windows highlighted in Figs.~\ref{fig:Experiment_TE_PBS}\textbf{c, f} and \ref{fig:Experiment_TM_PBS}\textbf{c, f}.
Experimentally, near-infrared images were acquired with an NIR camera and a 20x magnification objective lens.
Because the lattice footprint is large, we recorded out-of-plane scattered light from successive rings and
stitched the fields of view to visualize the full edge path. Background in waveguide-free regions was
suppressed using a Sobel operator \cite{gonzalez2009digital}.

\begin{figure}
    \centering
    \includegraphics[width=1\linewidth]{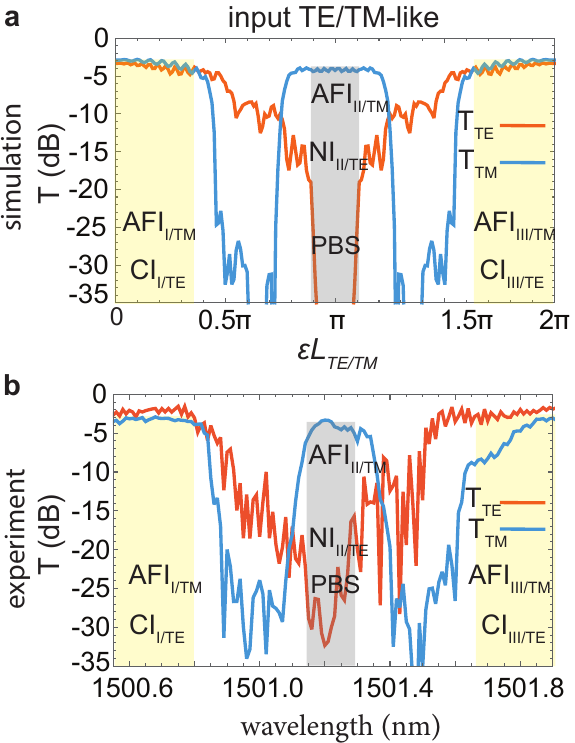}
    \caption{\textbf{Polarization-independent TPI.} \textbf{a} Simulated transmission (Port T) for TE- (orange trace) and reflection (R) for TM-polarized (blue trace) light. For TE- and TM-polarized light, the coupling powers between the microrings are  $\kappa_\text{{TE}}^2=70\%$ ($\theta_\text{{TE}}=0.35\pi$) and $\kappa_\text{{TM}}^2=87\%$ ($\theta_\text{{TM}}=0.62\pi$) respectively. \textbf{b} Measured transmitted power from port \textbf{T} for TE-like and TM-like input light. Simulated (measured) transmission is normalized to the maximum simulated (experimentally measured) power from ports \textbf{T} and \textbf{R}. The gray background shows the PBS region of operation. The yellow background represents the edge states which are independent of polarization. }
    \label{fig:Experiment_TE/TM}
\end{figure}

In addition to spectral regions where the lattice functions as a PBS, there exist wavelengths at which it supports topological band gaps for both polarizations, yielding polarization-independent edge states. Around \(\lambda \approx 1500~\mathrm{nm}\), where the TE–TM phase offset in the band diagrams is near zero, simulations (Fig.~\ref{fig:Experiment_TE/TM}\textbf{a}) show overlapping TE and TM edge states in the first and third topological band gaps (highlighted in yellow), while the middle band gap (gray) operates as a PBS. Corresponding measurements (Fig.~\ref{fig:Experiment_TE/TM}\textbf{b}), obtained by injecting TE-like and TM-like inputs separately, confirm that the first and third band gaps support polarization-independent topological edge transport. Most of the polarization-independent edge states are realized in static topological systems \cite{hu2023observation}; thus, this is the first experimental demonstration of a polarization-independent topological photonic insulator in Floquet topological photonic systems.

\subsection*{Robustness Against Defects}
In the proposed PBS, only the transmitted polarization is topologically protected, whereas the orthogonal polarization is reflected due to the topologically trivial (normal) bandgap and lacks such protection. This topological protection of the transmitted light substantially improves the overall performance of the PBS by strongly suppressing backscattering from defects. In contrast to conventional non-topological PBS designs—such as those based on polarization-selective bandgaps formed by periodic structures~\cite{shi2010experimental,xu2019compact, schonbrun2006polarization}, where imperfections scatter light and redirect it to the undesired port—the proposed topological PBS supports a unidirectional edge state that eliminates backscattering for the transmitted polarization. Although the structure supports both forward and backward propagating edge states, these states do not couple under typical perturbations. Coupling between forward and backward modes requires specific defects capable of flipping the pseudo-spin degree of freedom, commonly referred to as “magnetic disorder” \cite{hafezi2011robust, gao2016probing}. Fabrication-induced imperfections, such as surface roughness, generally induce only weak pseudo-spin mixing. Experimental studies of waveguides and coupled microring resonators consistently demonstrate that this mixing remains negligible under realistic conditions \cite{melati2014real,ferrari2009disorder, ferrari2011penalty}. Beyond pseudo-spin-flipping defects, other perturbations—sometimes described as “backscattering-immune strongly dissipative defects” which can destroy the edge states—have also been examined. Both theoretical predictions and experimental results in topological photonic systems based on two-dimensional coupled microring resonator lattices confirm that edge states exhibit high immunity to backscattering from most common fabrication and design imperfections \cite{gao2016probing, hafezi2013imaging}.

\begin{figure*}
    \centering
    \includegraphics[width=1\linewidth]{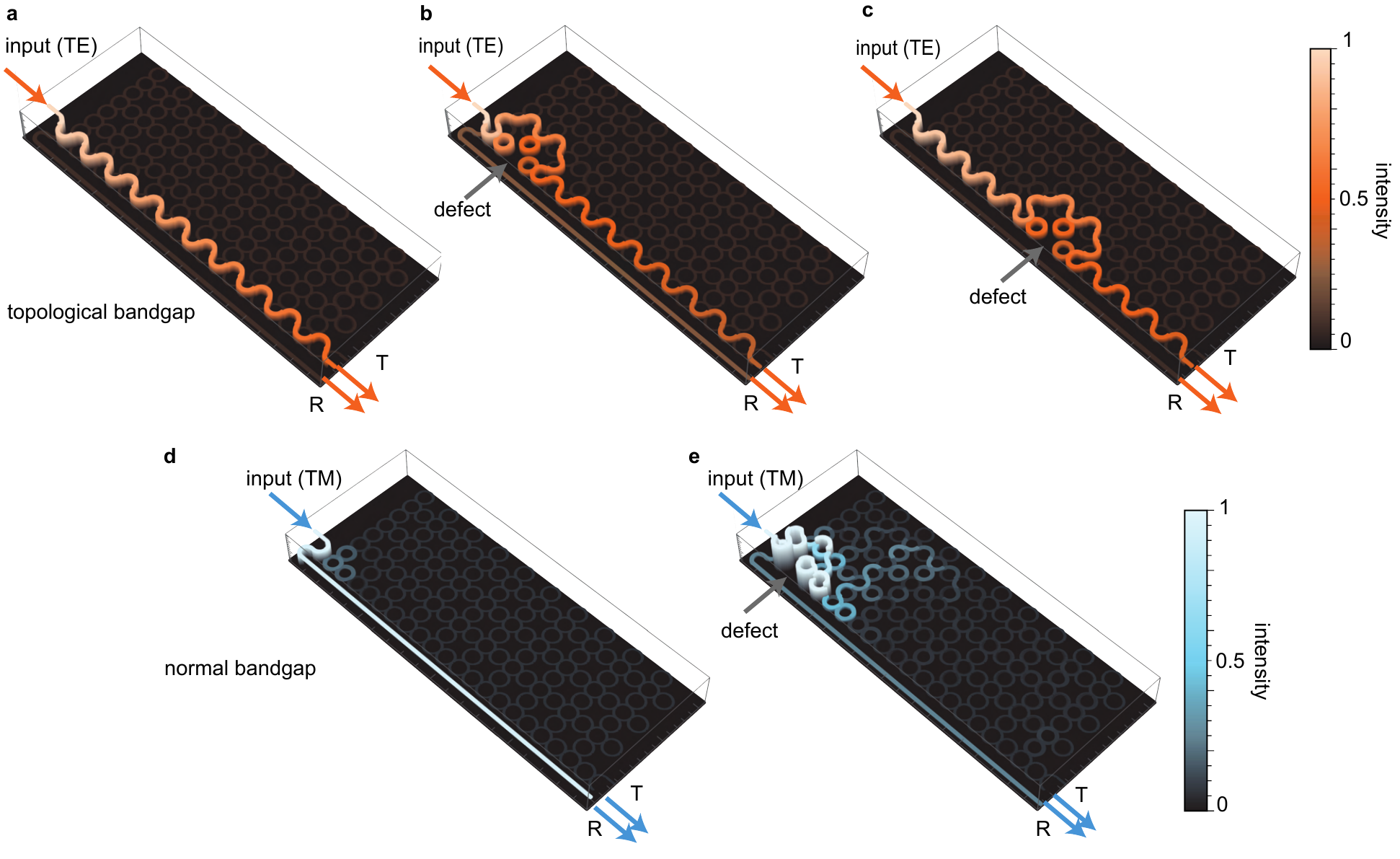}
    \caption{Simulated intensity distributions for the TE-protected topological PBS, illustrating the effect of defects on edge-state propagation and reflection.   (a–c) ISpatial intensity distributions of the TE-polarized edge state at $\epsilon L_{TE} =  \pi$ (corresponding  to $\lambda = 1625.55$ nm). \textbf{a} Defect-free lattice, showing TE-polarized edge state propagation along the edge.   \textbf{b} Lattice with a defect near the input port, resulting in partial light leakage into the reflection port.   \textbf{c} Lattice with a defect far from both input and output ports, exhibiting no backscattering and robust edge-state transmission.   \textbf{d, e} Field intensity distributions of TM-polarized light reflected from the conventional (normal) bandgap at $\epsilon L_{TM} = 0.32\pi$ (corresponding to $\lambda = 1624.7$ nm). \textbf{d} Defect-free lattice.   \textbf{e} Lattice with a defect near the input port, showing increased reflection due to local resonances near the defect site.} 
    \label{fig:robust_sim_3D_defect}
\end{figure*}

Here, we first employ analytical methods to examine the influence of defects on the transmission and reflection of TE- and TM-polarized light across different operating regimes of the proposed lattice. The robustness of the PER at each output port is also evaluated. Defects were introduced by removing a single ring from the edge-state path at various locations: near the input port, near the output port, and in the central region far from both ports. Representative examples of these defect configurations are illustrated in Fig.~\ref{fig:robust_sim_3D_defect}.

\begin{figure}[h!]
    \centering
    \includegraphics[width=1\linewidth]{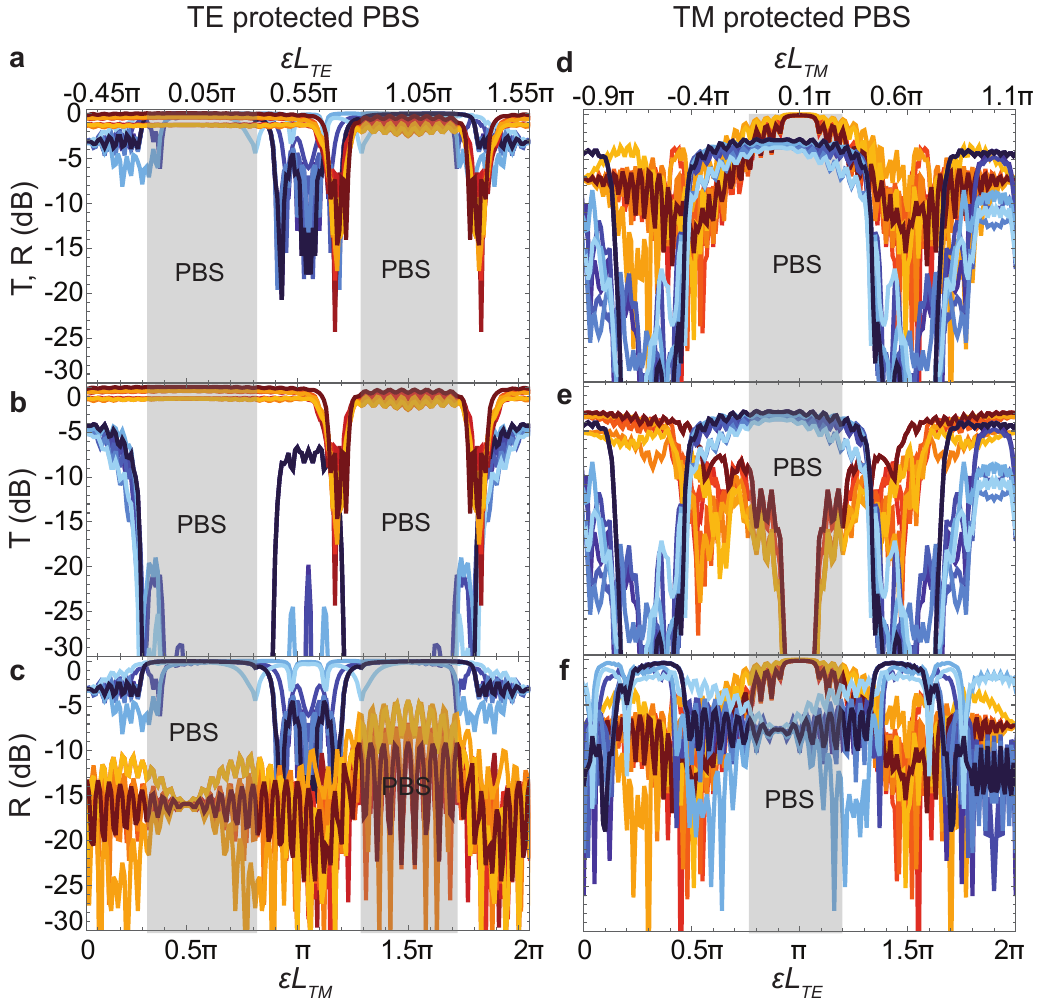}
    \caption{Simulated results of nine PBS: one defect-free lattice (dark orange for TE-polarized light, dark blue for TM-polarized light) and eight defect positions along the bottom edge path. Color shading indicates defect proximity to the input and output ports, with lighter shades corresponding to defects closer to the ports and darker shades to those farther away.   \textbf{a-c} show the PBS performance and the PER for the \textbf{T} and \textbf{R} ports at the longer-wavelength regime for the TE-protected topological PBS. Coupling strengths are taken from experimental measurements: $\kappa_\text{{TE}}^2 = 98.5 \%$ ($\theta_\text{{TE}} = 0.46\pi$) for TE-polarized light and $\kappa_\text{{TM}}^2 = 53 \%$ ($\theta_\text{{TM}} = 0.74\pi$) for TM-polarized light. \textbf{d-f} show the PBS performance and the PER for the \textbf{T} and \textbf{R} ports at the shorter-wavelength regime for the TM-protected topological PBS. Coupling strengths are taken from experimental measurements: $\kappa_\text{{TE}}^2=82\%$ ($\theta_\text{{TE}}=0.36\pi$) and $\kappa_\text{{TM}}^2=84\%$ ($\theta_\text{{TM}}=0.63\pi$)}  
    \label{fig:robust_sim_one_ring}
\end{figure}

Figures~\ref{fig:robust_sim_one_ring}\textbf{a, d}  present simulated transmission and reflection spectra of the PBS in the presence of defects at high and low operating wavelengths, respectively, using coupling strengths extracted from experimental data. Dark orange and dark blue lines correspond to TE- and TM-polarized light in the defect-free case, while lighter-to-darker shades of orange and blue represent results for eight defect positions, with color intensity scaled by defect proximity to the input port (lighter shades for defects closer to the input, darker for those farther away). As shown in Fig.~\ref{fig:robust_sim_one_ring}\textbf{a, d}, transmission of the topologically protected modes remains highly robust. Ring removal in the first or second unit cell adjacent to the input or output ports induces losses of up to 2 dB for the TE mode and 3.4 dB for the TM mode. In contrast, the unprotected reflected modes exhibit greater sensitivity: defects cause drops of up to 6.8 dB in the reflected TM mode (light blue curves in Fig.~\ref{fig:robust_sim_one_ring}\textbf{a}),  while changes in the reflected TE mode Fig.~\ref{fig:robust_sim_one_ring}\textbf{d}) remain negligible. 

To identify the origin of these losses and assess their impact on PER, port \textbf{T} (Fig.~\ref{fig:robust_sim_one_ring}\textbf{b, e}) and port \textbf{R} (Fig.~\ref{fig:robust_sim_one_ring}\textbf{c, f}) are analyzed separately. For defects far from the input port, reflection at port \textbf{R} remains virtually unchanged, indicating that the observed transmission loss at port \textbf{T} arises primarily from propagation loss due to path detouring around the defect. Only for defects in very close proximity to the input port does reflection increase by a few dB. Since identical defects placed farther from the input produce no such increase, this effect is not attributable to backscattering from forward–backward edge-state coupling. Rather, it arises from the defect-induced detour occurring near the input, which allows a portion of the light to couple into the reflection port. 

To further illustrate the topological robustness, we examine the intensity distributions of polarized light in both topologically nontrivial and trivial bandgaps for the TE-protected PBS.  At $\epsilon L_{TE} = \pi$ (topologically nontrivial bandgap), intensity distributions are shown for the defect-free lattice (Fig. ~\ref{fig:robust_sim_3D_defect}\textbf{a}), with a defect close to the input port (Fig.~\ref{fig:robust_sim_3D_defect}\textbf{b}), and with a defect far from the input port (Fig.~\ref{fig:robust_sim_3D_defect}\textbf{c}). As anticipated, the near-input defect causes a small fraction of light to enter the reflection port, whereas the remote defect produces no backscattering—confirming the proposed PBS's immunity to backscattering. Suppression of backscattering preserves the polarization extinction ratio under defect perturbations, thereby enhancing overall device performance.  Figures~\ref{fig:robust_sim_3D_defect}\textbf{d, e} present intensity distributions for TM-polarized light at $\epsilon L_{TM} = 0.32 \pi$ (topologically trivial bandgap), before and after introducing a defect near the input port, where a 6.8 dB loss appears at port \textbf{R}. This loss originates from local resonances induced near the input. As evident in   Fig.~\ref{fig:robust_sim_one_ring}\textbf{a}, defects cause only minor changes in the reflected polarized light across most of the spectrum (except at a few points). These observations underscore that a 2D lattice structure provides substantially greater defect tolerance than conventional 1D bandgap-based PBS designs.

\begin{figure}
    \centering
    \includegraphics[width=1\linewidth]{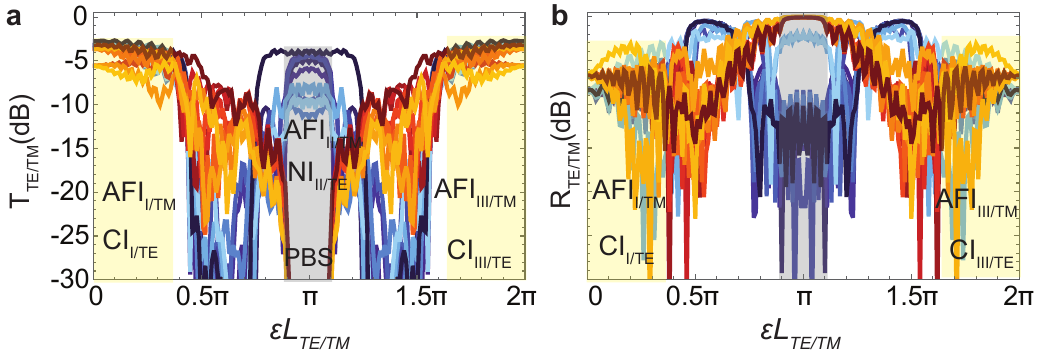}
    \caption{\textbf{a, b} Simulated transmission and reflection spectra for the dual-polarization topological lattice in the presence of defects. The coupling powers are derived from experimental measurements:  $\kappa_\text{{TE}}^2=70\%$ ($\theta_\text{{TE}}=0.35\pi$) and $\kappa_\text{{TM}}^2=87\%$ ($\theta_\text{{TM}}=0.62\pi$). Yellow-highlighted regions indicate the spectral range supporting dual-polarized edge states (robust transmission for both TE and TM polarizations), while gray-highlighted regions denote the PBS-operating regime where only TM polarization is topologically protected. }  
    \label{fig:robust_sim_one_ring+_dual_polarization}
\end{figure}

Analogous simulations were performed for dual-polarization edge states. Transmission (port \textbf{T}) and reflection (port \textbf{R}) spectra are shown in Fig.~\ref{fig:robust_sim_one_ring+_dual_polarization}\textbf{a, b}, respectively. Yellow-highlighted regions denote topological bandgaps supporting both TE- and TM-polarized light, while gray regions indicate the topological (nontrivial and trivial) bandgaps exploited for PBS operation, which protect TM-polarized light. Transmission results demonstrate robustness of dual-polarized edge states against defects, except for TE-polarized light, where weaker  coupling ($\kappa^2=70\%$) results in losses of approximately 2.4 dB at port \textbf{T} for defects near the input and output ports.   Reflection spectra reveal that, near the TE resonance  ($\epsilon L_{TE} = 2 \pi$), losses arise predominantly from propagation effects due to path detouring; at other frequencies, a portion of the light couples into port \textbf{R}. In the central bandgap for TM-polarized light, defects near the ports induce transmission losses of up to 7.5 dB, with reflection spectra confirming significant light transfer to port \textbf{R} for near-input defects.

These results demonstrate that defect-induced losses in edge states depend on three primary factors. First, defect location: losses occur only when the defect is in close proximity to the input port. Second, coupling strength: for strong coupling ($\kappa^2=98.5 \%$, as in Fig.~\ref{fig:robust_sim_one_ring}\textbf{a}), losses remain below 2 dB, whereas weaker coupling ($\kappa^2=87 \%$ in Fig.~\ref{fig:robust_sim_one_ring+_dual_polarization}\textbf{a}) increases losses to as much as 7.5 dB. Third, bandwidth of topological bandgap: as shown in Fig.~\ref{fig:robust_sim_one_ring+_dual_polarization}\textbf{a}, losses in the wider bandgaps (I and III) are significantly lower than in the narrower bandgap II. These findings indicate that achieving a topological PBS with negligible defect-induced loss at the output ports requires designing the lattice for strong coupling between microrings.

\begin{figure*}
    \centering
    \includegraphics[width=1\linewidth]{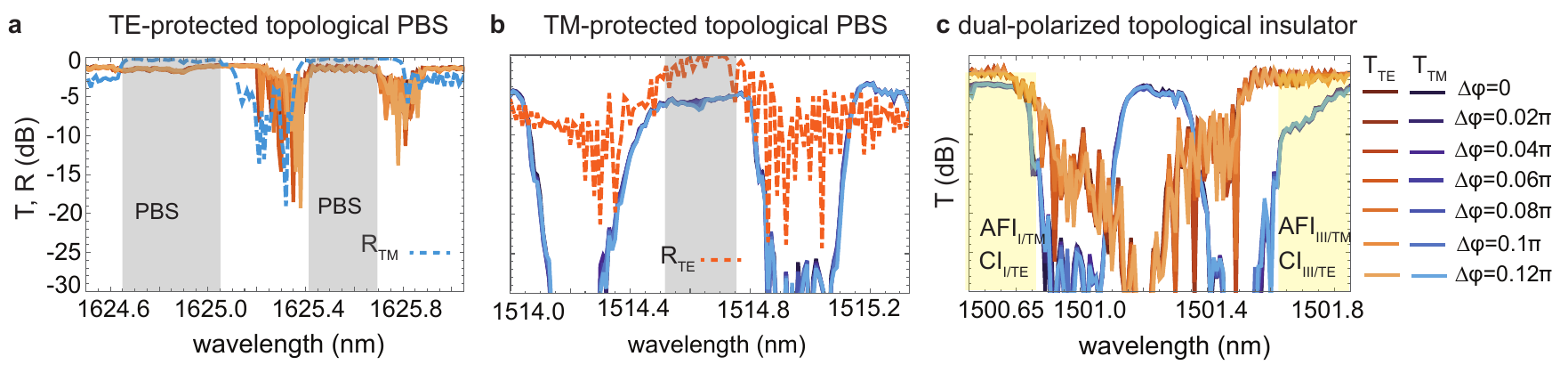}
    \caption{Experimental demonstration of robustness against a local defect in the proposed topological PBS lattice. The defect is induced thermo-optically by heating a single microring along the edge path, which locally modifies its refractive index and round-trip phase $\phi$ without causing significant backscattering.   
     \textbf{a-c} Different shades of solid orange and blue lines represent the \textbf{T} port, respectively, for TE- and TM polarized light after applying various heating powers for the TE/TM-protected topological PBS, and dual polarized edge states. Dashed orange and blue lines, respectively, indicate the TE and TM polarized light reflected into port \textbf{R}.} 
    \label{fig:robust_exp}
\end{figure*}

\noindent\textbf{Experimental Demonstration of Robustness.} To experimentally validate the robustness of the proposed lattice, a defect was introduced by locally modifying the refractive index of a microring on the edge path via the thermo-optic effect leading to change the round-trip phase $\phi_{rt}$ of the microring. Figure~\ref{fig:robust_exp}\textbf{a-c}  display transmission and reflection spectra measured at three distinct operating wavelengths. The results confirm that transmission through the edge states remains robust against the induced defect. Higher heating powers could not be applied, as they risked damaging the TiW alloy heater fabricated on top of the microring.

\section*{Discussion}
We have introduced a CMOS-compatible, defect-tolerant polarization beam splitter that leverages
a Floquet topological microring lattice to separate TE and TM by exploiting polarization-dependent band topology.
By engineering adjacent trivial and nontrivial band gaps, the lattice provides distinct topological environments for the two
polarizations, ensuring that one polarization couples to a protected edge state while the other is rejected. This mechanism
provides intrinsic resilience to fabrication variations and localized defects, in contrast to conventional PBS designs that rely
on tight geometric tolerances. 
In addition, at selected wavelengths, the same lattice supports
simultaneous topological band gaps for both polarizations, yielding polarization-independent edge transport. We note that previous proposals for topological polarization beam splitters based on photonic crystals have remained at the theoretical level \cite{su2023compact, he2024topological, he2021topologically, he2022topological, li2025t, deng2025ultrahigh}. These approaches are based on static topological phases, which typically support edge states over relatively narrow spectral ranges, in contrast to Floquet systems that enable operation across multiple free spectral ranges. In addition, these designs rely on a different mechanism for polarization separation. Specifically, they require two topological lattices that each support dual-polarization edge states, with TE and TM modes separated through engineered coupling at the interface between the lattices. This approach demands precise control of both polarization-dependent band topology and inter-lattice coupling. Such requirements impose strict constraints on fabrication and increase the complexity of integration with other photonic components. In contrast, the present work achieves polarization-selective routing within a single Floquet microring lattice, without relying on interface engineering, and is directly compatible with standard CMOS-based platforms.

The per-FSR bandwidth is presently limited by the small FSR of the large-radius SiN rings chosen to suppress bend loss. This constraint is technological rather than topological: moving to a higher-index platform enables smaller radii and larger FSR at comparable loss without changing the AFI phase of the Floquet operator. As shown in Table~\ref{table1}, a design-verified 220 nm silicon variant with R=5  $\mu$m yields an FSR of 5.9 nm at 1550 nm (bend loss $<0.5$ dB/cm) and improves the extinction ratio owing to stronger coupling power between microrings. 
A high-contrast SiN option (height 400 nm, R=25 $\mu$m) provides an FSR of 6.6 nm with similar loss. Thus, bandwidth is set by ring dispersion—not by the topological mechanism—and can be expanded without altering the phase diagram. The measured extinction ratio of 10 to 20 dB at high wavelengths and 7 to 25 dB for lower wavelengths can be further improved by strengthening bus-to-lattice coupling and tightening coupling power uniformity.

Beyond a fabrication-tolerant PBS, a Floquet topological microring lattice with
polarization-dependent band topology has immediate applications in both quantum and classical domains. In classical photonics, it supports polarization
de/multiplexing and polarization-mode dispersion–tolerant routing for coherent transceivers and WDM backbones, while
its FSR-spanning behavior facilitates broadband polarization management in comb-based systems (dual-comb spectroscopy and
on-chip metrology). In quantum photonics, robust TE/TM sorting at fixed wavelengths streamlines polarization-encoded qubit
preparation and analysis and reduces crosstalk by confining protected channels to designed boundaries. For sensing and
LiDAR-style ranging, polarization-selective edge transport suppresses backscattering-induced baseline drift and stabilizes
the state-of-polarization readout over temperature and aging. Because the band-to-port mapping naturally varies with wavelength,
the same architecture can also function as a compact, passive polarization switch for wavelength-agile sources. Furthermore, our topological PBS pave the way for more complex topological PBS in that both TE and TM polarized light can be topologically protected using topological heterostructures based on Floquet TPIs, which provide a 2D robust and insensitive environment for programmable TE and TM polarized devices, specifically for interferometers measuring polarization-entangled quantum states.

\textbf{Acknowledgments}
S.B.\ acknowledges funding by the Natural Sciences and Engineering Research Council of Canada (NSERC) through its Discovery Grant, funding and advisory support provided by Alberta Innovates through the Accelerating Innovations into CarE (AICE) – Concepts Program, and support from Alberta Innovates and NSERC through Advance Grant.

\textbf{Data availability}
The raw data that support the findings of this study are available from the corresponding authors upon reasonable request.

\textbf{Contributions}
Experiments and data analysis were carried out by S.A and A.K. The project was supervised and directed by S.B and the study was conceived by S.B and S.A

\section*{Methods}

\noindent\textbf{Device fabrication.}
To experimentally demonstrate a topological PBS, we fabricated a TPI microring lattice with $4\times 10$ unit cells on a SiN-on-insulator chip (fabricated by Applied Nanotools Inc.). The SiN device layer is 400~nm thick and is clad by 4.5~\(\mu\)m (lower) and 3~\(\mu\)m (upper) SiO$_2$. The lattice comprises square microrings formed from 1.6~\(\mu\)m-wide waveguides with 150~nm gap between adjacent microrings. To reduce corner loss associated with the relatively low effective index ($\sim 1.7$), each square ring corner was rounded with a 120~\(\mu\)m radius.

\medskip
\noindent\textbf{Design and numerical modeling.}
We used Lumerical MODE (mode solver) to determine the side length of the square resonators and to extract the polarization-dependent coupling strength per unit length, $\theta/\ell_{\mathrm{TE/TM}}$, where $\ell_{\mathrm{TE/TM}}$ denotes the effective coupling length for the fundamental TE and TM modes. Both the ring side length and the corner-rounding radius contribute to the effective coupling. Based on simulation and corroborating measurements, a 120~\(\mu\)m corner radius increases the effective coupling length for both TE and TM modes.
Balancing these effects, we selected a ring side length of 38~\(\mu\)m, which yields polarization-dependent coupling dispersion: TM-protected topological PBS operation at shorter wavelengths and TE-protected operation at longer wavelengths.

\begin{figure}[h!]
    \centering
    \includegraphics[width=1\linewidth]{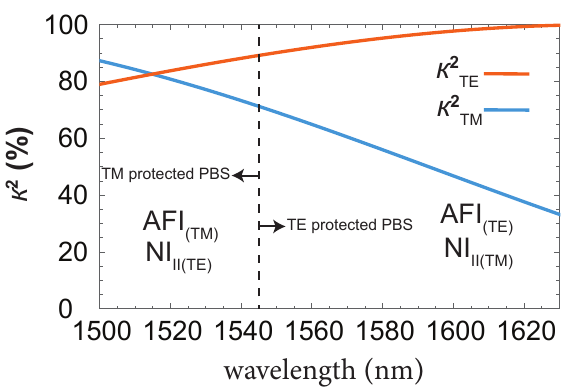}
    \caption{The simulated coupling power between microrings for TE-polarized light (orange line) and TM-polarized light (blue line) is obtained using Lumerical mode-solution simulation. This simulation considers the coupling length of two couplers with straight waveguides of 38 $\mu$m length and rounded corners with a radius of 120 $\mu$m. The dashed line shows the wavelength at which the central band gap closes to create a phase transition for both TE- and TM-polarized light. The left side of this figure operates  as a TM-protected PBS, and the right side operates as a TE-protected PBS.}
    \label{fig:coupling_sim}
\end{figure}

\noindent\textbf{Topological phase mapping and coupling dispersion.}
Figure~\ref{fig:coupling_sim} shows the simulated percentage of power coupled between adjacent square microrings for TE and TM modes. The wavelength dependence of the coupling leads to distinct topological behaviors for the two polarizations (see Ref.~\cite{afzal2018topological}). For TE, at shorter wavelengths, the energy-band diagram exhibits two AFI band gaps and one NI band gap within each FSR; near 1545~nm, the NI band gap closes, and an AFI band gap opens as the wavelength increases. For TM, the process is reversed: the AFI band gap closes around 1545~nm and an NI band gap opens at longer wavelengths. Although phase transitions from NI to AFI band gaps have been analyzed previously~\cite{afzal2018topological}, our simulations indicate that the coupling angle at which the transition occurs is shifted relative to the ideal 2D phase map when the lattice is finite. Energy-band calculations of semi-infinite lattices truncated along one direction, with varying numbers of unit cells, show that fewer unit cells shift the phase transition to higher coupling power. In addition, the TE and TM transitions occur at different coupling powers in our design (about 90\% for TE and 70\% for TM). This difference originates from wavelength-dependent coupling dispersion rather than polarization itself: for lattices where coupling power decreases with increasing wavelength, the transition occurs at a lower coupling power (i.e., a smaller coupling angle). Accounting for lattice size and coupling dispersion is, thus, essential when designing a 2D microring topological PBS. Exploiting the opposite coupling dispersion for TE and TM enables a TPI PBS that operates in two regimes: TM-polarized light is topologically protected at shorter wavelengths, while TE-polarized light is protected at longer wavelengths (highlighted in Fig.~\ref{fig:coupling_sim}).

\begin{figure}
    \centering
    \includegraphics[width=1\linewidth]{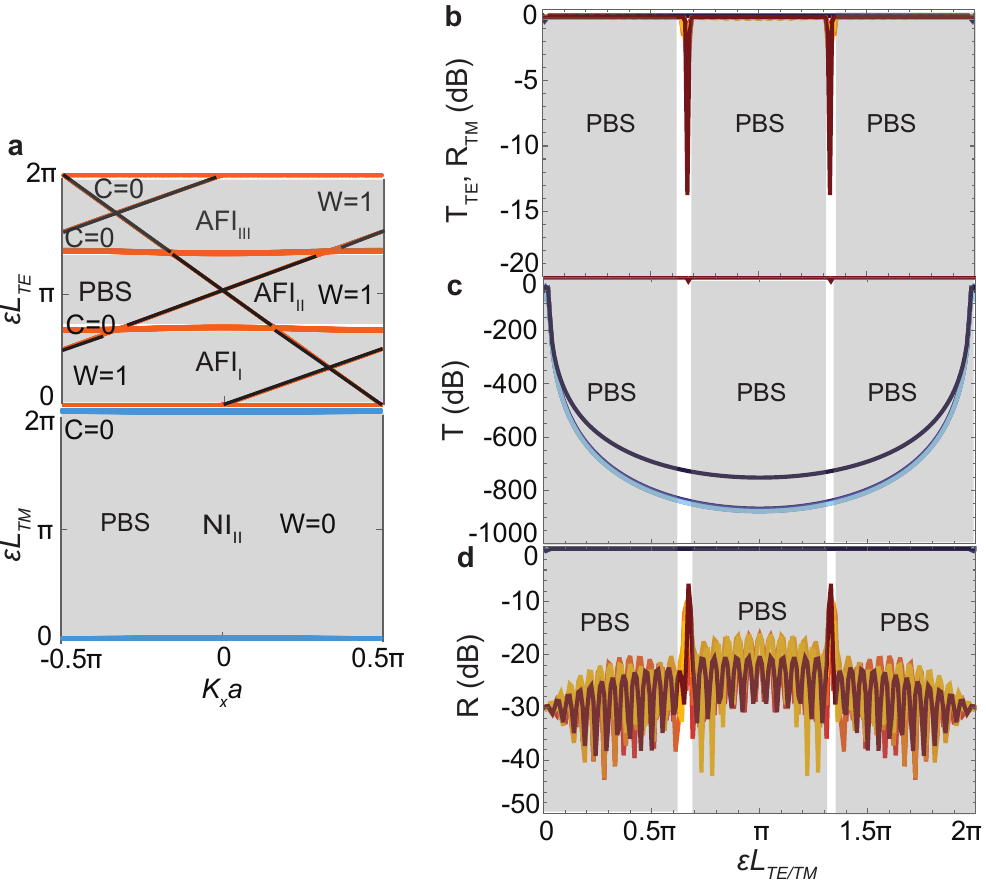}
    \caption{Suggested design for a high-performance TE-protected topological PBS with wide operational bandwidth and enhanced PER.   \textbf{a} Band diagram of the proposed lattice, engineered with strong TE coupling  ($\theta=0.49\pi$, $\kappa ^2>99\%$) and weak TM coupling  ($\theta=0.01\pi$). Gray-highlighted regions indicate the spectral ranges suitable for PBS operation (topological protection of TE-polarized light).   \textbf{b-d} PBS response, \textbf{T} and \textbf{R} ports spectra for the TE-protected PBS across nine configurations: one defect-free lattice (dark orange lines for TE-polarized light, dark blue lines for TM-polarized light) and eight defect positions along the edge path. Lighter-to-darker shades of orange and blue represent defect proximity to the input/output ports (lighter shades for defects closer to the ports, darker for those farther away). The results demonstrate negligible transmission loss and highly stable PER even in the presence of defects.} 
    \label{fig:robust_sim_one_ring_suggestiom}
\end{figure}

\noindent\textbf{Design Improvements for Higher Performance.}
The proposed lattice serves as a straightforward example of a topological photonic structure capable of operating in three distinct modes across different wavelengths: TE-protected PBS, TM-protected PBS, and dual-polarization edge states. Higher-performance PBS designs can be achieved by tailoring the lattice to exhibit strong coupling between microrings for one polarization while maintaining weaker coupling for the orthogonal polarization.

One such example of a TE-protected topological PBS is presented in Fig.~\ref{fig:robust_sim_one_ring_suggestiom} for the lattices without and with difects, simulated analytically with coupling angles of  $\theta=0.49\pi$ for TE-polarized light and  $\theta=0.01\pi$ for TM-polarized light  (Fig.~\ref{fig:robust_sim_one_ring_suggestiom}\textbf{a}). In the simulated, the input and ouput coupling strength matches the  coupling within the lattice. The results demonstrate negligible transmission loss in the presence of defects, even when located near the input and output ports. PER in each port is significantly enhanced and remains stable under most defect perturbations, with only those defects close to the input port causing a reduction of approximately 4 dB in the central bandgap region. As shown in Fig.~\ref{fig:robust_sim_one_ring_suggestiom}\textbf{c, d}, the PER for the topologically protected mode exceeds 30 dB and reaches up to 750 dB near the center of one FSR, whereas the PER at the unprotected reflection port \textbf{R} is limited to 20–30 dB. Given the negligible propagation loss in the lattice, PER values at both \textbf{T}  and \textbf{R} ports are primarily constrained by leakage at the input coupler.  Consequently, the observed 30 dB PER limit arises from expressions of the form $10\times log(\kappa_{in, TE}^2/\ A \kappa_{in, TM}^2)$ for port \textbf{T} and $10 \times log(1-\kappa_{in, TM}^2/1-\kappa_{in, TE}^2)$ for port \textbf{R}, where \textbf{\textit{A}} accounts for bulk-mode losses, and $\kappa_{in, TE/TM}^2$ denotes the input/output coupling powers for each polarization. For the protected \textbf{T} port (Fig.~\ref{fig:robust_sim_one_ring_suggestiom}\textbf{c}), this limitation is confined to a narrow spectral region near the resonance  ($\epsilon L_{TE/TM} = 2 \pi$), while leakage from the conventional bandgap remains negligible and scales with lattice size. At the unprotected \textbf{R} port  (Fig.~\ref{fig:robust_sim_one_ring_suggestiom}\textbf{d}), PER is further reduced by a small fraction of edge-state light that cannot couple to the output port, forcing it to detour around the lattice structure and enter the reflection port. This constraint can be mitigated through optimized input and output coupler designs.

\begin{table*}[ht!]
  \centering 
  \caption{ Comparison of FSR and estimated total loss for topological PBS implemented in silicon (Si) and silicon nitride (SiN) microring lattices with different bending radii. Si waveguides have cross-sectional dimensions of 200 nm (height) $\times$ 450 nm (width); SiN waveguides have 400 nm height and widths of 1 $\mu$m or 1.6 $\mu$m. The gap between microrings is considered to be fixed at 150 nm across all designs. $^{a}$ Total loss values are estimated by considering only the bending loss in the topological edge-state path across 4$\times$10 unit cells. Propagation loss and radiation loss are assumed negligible in these calculations.  Simulations indicate that total loss can be further reduced by optimizing the lattice size (e.g., using a more compact $4\times4$ unit-cell PBS design) to shorten the optical path length.} 
\begin{tabular}{|l|c|c|c|c|}
\hline
\textbf{Material and the Radius of rounded-corners} & \textbf{coupling length ($\mu$m)} &\textbf{FSR (nm) } &  \textbf{total  loss$^{a}$ dB}  \\ 
\hline
Si-TE protected (simulation, $r=5~\mu$m)  & 15.8 & 5.9  &   -0.4284  \\
\hline
Si-TM protected  (simulation, $r=5~\mu$m)& 4.5 & 13.5 &   -14.3556 \\
\hline
Si-TE protected  (simulation, $r=10~\mu$m)& 15.8 & 4.4 &    -0.1092
  \\
\hline
Si-TM protected  (simulation, $r=10~\mu$m)& 4.5 & 8.2 &    -2.0328 \\
\hline
Si-TE protected  (simulation, $r=15~\mu$m)& 15.8 & 3.5 &    -0.0504  \\
\hline
Si-TM protected  (simulation, $r=15~\mu$m)& 4.5 & 5.9 &  -0.8064 
 \\
\hline
SiN (width=1 $\mu$m)-TE protected  (experiment, $r=25~\mu$m)& 10 & 6.6& -0.7476
  \\
\hline
SiN (width=1 $\mu$m)-TM protected  (experiment, $r=25~\mu$m)& 10 & 6.1 &   -1.5624  \\
\hline
SiN (width=1.6 $\mu$m)-TE protected  (simulation, $r=25~\mu$m)& 38 & 4.1 &    -10.0879  \\
\hline
SiN (width=1.6 $\mu$m)-TM protected  (simulation, $r=25~\mu$m)& 38 & 3.8 &    -13.5800  \\
\hline
SiN (width=1.6 $\mu$m)-TE protected  (this work, $r=120~\mu$m)& 38 & 1.4 &  -0.4536
  \\
\hline
SiN (width=1.6 $\mu$m)-TM protected  (this work, $r=120~\mu$m)& 38 & 1.3 &  -0.6216  \\
\hline
\end{tabular}
\label{table1}
\end{table*}

In addition, as indicated by the gray regions in Fig.~\ref{fig:robust_sim_one_ring_suggestiom}\textbf{a}, approximately 94\% of one FSR supports PBS operation, with only 6\% forming a dead zone. To achieve even wider bandwidth, silicon (Si) microrings can replace silicon nitride (SiN), leveraging the larger FSR of Si while maintaining low propagation losses. We simulated a 4$\times$10 unit-cell topological photonic insulator  based on Si microrings (waveguide width 450 nm, gap 150 nm, bending radii of 5, 10, and 15 $\mu$m), assuming coupling strengths comparable to those in the suggested lattice of Fig~\ref{fig:robust_sim_one_ring_suggestiom}. 

Table~\ref{table1} summarizes the estimated FSR and total loss (dominated by bending loss, with propagation and radiation losses assumed negligible). For a TE-protected PBS with 5 $\mu$m bending radii, an FSR of 5.9 nm is attainable, with total loss for an edge state traversing 10 unit cells below 0.5 dB. For a TM-protected PBS with 10  $\mu$m bending radii, the FSR increases to 8.2 nm,  however with  approximately 2 dB total loss. Larger bending radii further reduce losses. We also fabricated another SiN-based TPI lattice (waveguide width 1  $\mu$m, gap = 150 nm, bending radius 25 $\mu$m), achieving FSR greater than 6 nm and total loss less than 1 dB for the TE-protected PBS (with estimated less than 2 dB for TM-protected operation). This design performs well for TE protection PBS but is limited by weak TM coupling, which limits TM-protected performance. Simulations further explore the impact of same bending radii (25 $\mu$m) combined with wider SiN waveguides (1.6 $\mu$m width). Importantly, the PBS bandwidth is not restricted to a single FSR and can span multiple FSRs, depending on waveguide dispersion over a broad wavelength range.

Finally, to realize a PBS that topologically protects both TE- and TM-polarized light simultaneously, a heterostructure approach can be employed, where one side of the PBS protects TE-polarized light (reflecting TM) and the opposite side protects TM-polarized light (reflecting TE). The primary challenge in such a heterostructure PBS lies in designing an efficient input coupler capable of selectively exciting the protected edge states on each side.

\bibliography{refs}
\end{document}